%% file: main.tex
\documentclass[10pt,twocolumn,letterpaper]{article}

\usepackage[pagenumbers]{cvpr} 
\newcommand{\ceil}[1]{\left\lceil #1 \right\rceil}
\usepackage{multirow}
\usepackage{makecell}

\definecolor{cvprblue}{rgb}{0.21,0.49,0.74}
\usepackage[pagebackref,breaklinks,colorlinks,allcolors=cvprblue]{hyperref}

\def\paperID{16514} 
\def\confName{CVPR}
\def\confYear{2025}

\title{Stochastic BIQA: Median Randomized Smoothing for Certified Blind Image Quality Assessment}

\author {
    Ekaterina Shumitskaya\textsuperscript{\rm 2, 1, 3},
    Mikhail Pautov\textsuperscript{\rm 5,2},
    Dmitriy Vatolin\textsuperscript{\rm 1,2,3}
    Anastasia Antsiferova\textsuperscript{\rm 1,2,4}\\
    \textsuperscript{\rm 1}MSU Institute for Artificial Intelligence\\
    \textsuperscript{\rm 2}ISP RAS Research Center for Trusted Artificial Intelligence\\
    \textsuperscript{\rm 3}Lomonosov Moscow State University\\
    \textsuperscript{\rm 4}Laboratory of Innovative Technologies for Processing Video Content, Innopolis University\\
    \textsuperscript{\rm 5}Artificial Intelligence Research Institute\\
}

\begin{document}

\newtheorem{definition}{Definition}[section]
\newtheorem{remark}{Remark}
\newtheorem{theorem}{Theorem}
\newtheorem{lemma}{Lemma}

\maketitle
\input{sec/0_abstract}   
\input{sec/1_intro}
\input{sec/2_formatting}
\input{sec/3_finalcopy}
{
    \small
    \bibliographystyle{ieeenat_fullname}
    \bibliography{main}
}


\end{document}

%% file: sec/0_abstract.tex
\begin{abstract}
Most modern No-Reference Image-Quality Assessment (NR-IQA) metrics are based on neural networks vulnerable to adversarial attacks. Attacks on such metrics lead to incorrect image/video quality predictions, which poses significant risks, especially in public benchmarks. Developers of image processing algorithms may unfairly increase the score of a target IQA metric without improving the actual quality of the adversarial image. Although some empirical defenses for IQA metrics were proposed, they do not provide theoretical guarantees and may be vulnerable to adaptive attacks. This work focuses on developing a provably robust no-reference IQA metric. Our method is based on Median Smoothing (MS) combined with an additional convolution denoiser with ranking loss to improve the SROCC and PLCC scores of the defended IQA metric. Compared with two prior methods on three datasets, our method exhibited superior SROCC and PLCC scores while maintaining comparable certified guarantees. We made the code available on GitHub: \textit{link is hidden for a blind review}.
\end{abstract}


%% file: sec/1_intro.tex
\section{Introduction}
\label{sec:intro}

\begin{figure*}[htb]
\begin{center}
\centerline{\includegraphics[width=\linewidth]{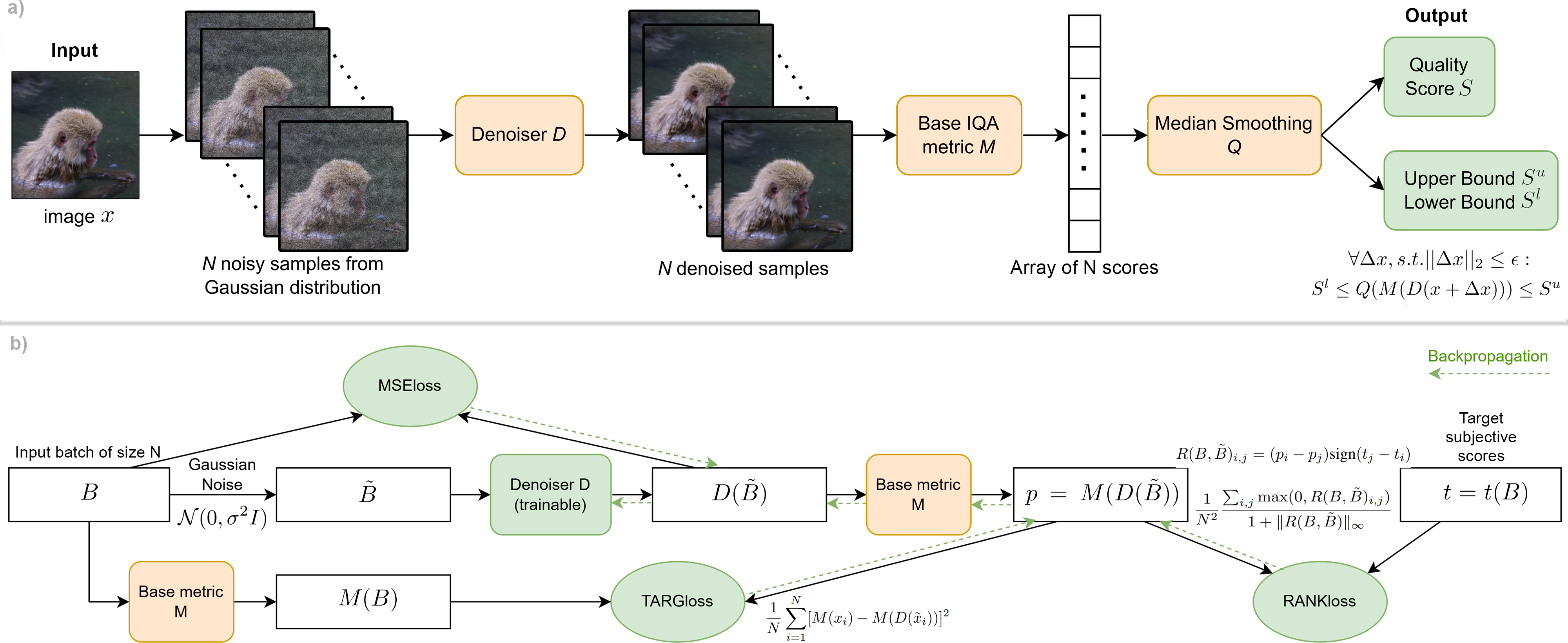}}
\caption{a) The overview of the proposed DMS-IQA method. b) Training scheme of the auxiliary denoiser.}
\label{fig:method_overview}
\end{center}
\end{figure*}

Image quality assessment (IQA) methods are widely used for quality control in video streaming services and public benchmarks. They are also used as loss functions when developing and optimizing video processing algorithms and in many other image/video processing applications. The goal of an IQA method is to estimate the perceptual quality of a given image, and the performance of these methods is illustrated by the correlation between the predictions and quality labels provided by assessors. Since most modern IQA metrics are based on neural networks, they are susceptible to synthetic perturbations of the input data, such as adversarial attacks \cite{goodfellow_explaining_2015, korhonen2022adversarial, antsiferova2024comparing, SHUMITSKAYA2024103913}. An adversary can apply image processing to artificially boost the score of an IQA metric without any actual improvement in the image's visual quality. This kind of attack can result in poor performance of video control services and unfair comparison results in public benchmarks\cite{antsiferova2024comparing}. Optimizing image processing algorithms using a vulnerable quality metric may significantly decrease the actual quality of processed images despite having high objective quality by the IQA metric. For example, training image- or video super-resolution methods with no-reference metrics in the objective function leads to visual artifacts \cite{kashkarov2024can}.

The vulnerability of IQA methods to adversarial attacks is currently an issue: many attacks have been proposed literature \cite{korhonen2022adversarial, SHUMITSKAYA2024103913, zhang2022perceptual, pmlr-v235-shumitskaya24a, siniukov2022applicability, zvezdakova2019hacking}, a benchmark of metrics' robustness~\cite{antsiferova2024comparing}, but almost no universal methods for IQA defenses exist. Some works focused on improving the robustness of particular metrics, such as LPIPS \cite{kettunen2019lpips, ghazanfari2023r}, or used standard purification defenses for verifying attack' performance under defense \cite{gushchin2024adversarial}. The primary challenge for the adversarial defense method for IQA is balancing the model accuracy and robustness trade-off \cite{gushchin2024guardians, antsiferova2024comparing}. Adversarial defenses fall into two main categories: empirical and provable defenses. The primary empirical defenses are adversarial training, which was adopted for IQA metrics \cite{technologies12110220, korhonen2022adversarial}, and adversarial purification methods that remove perturbations from input data. These methods do not provide mathematical guarantees: the defense can be inefficient against stronger adversaries. Also, compared to image classification models, the defense for an IQA metric must keep the visual quality of an image, which significantly complicates applying adversarial purification approaches such as blurring the image~\cite{gushchin2024adversarial}. In contrast, provable defenses provide guarantees for defended models, for example, to be robust against perturbations of particular norms. Only one work proposed a certified defense for IQA metrics via 1-Lipshitz neural networks \cite{ghazanfari2023lipsim}. However, the authors considered a full-reference setting for image quality assessment (when the IQA metric estimates the similarity between two images). Thus, the task of developing a certified no-reference IQA method, which estimates the quality by only one image, remains largely unexplored.

Our research aims to explore the potential of implementing randomized smoothing-based defenses to certify general-purpose no-reference (NR) metrics in blind IQA. This paper presents a defense technique --- Denoised Median Smoothing for IQA (DMS-IQA) --- that can be applied to any NR metric without imposing restrictions on the model's architecture. Additionally, the proposed defense does not require re-training the model itself. Still, it only involves the additional training of a small-scale U-Net denoiser, considering the relative distance between images in the batch during training. Figure \ref{fig:method_overview} shows an overview of the proposed DMS-IQA method. The proposed method guarantees that on some image $x$, any perturbation $\Delta x$ with an $l_2$ norm less than $\epsilon$, the IQA score is bounded and will remain between $S^l$ and $S^u$. We define the Certified Delta (CD) as the difference $CD = S^u - S^l$. Our main contributions are as follows:
\begin{itemize}
\item We propose a novel application of median randomized smoothing to create a certified defense for image quality metrics. To the best of our knowledge, this is the first work to propose a method for certified no-reference image quality assessment.
\item We propose to train the  denoiser to improve both the alignment in predictions of smoothed image quality assessment metric with its source counterpart and to reduce the deviation DMS-IQA's predictions from the subjective scores. 
\item We conduct comprehensive experiments assessing five no-reference metrics on three datasets to compare our method with prior approaches.
\item We demonstrate that the smoothed metric can be effectively used as a loss function for image-processing algorithms, using image denoising as an example.
\item We made our code available on GitHub: \textit{link is hidden for a blind review}.
\end{itemize}

%% file: sec/2_formatting.tex
\section{Related work}
\label{sec:formatting}

\subsection{Attacks and defenses on image quality metrics}
Image quality assessment (IQA) metrics can be divided into full-reference (FR) and no-reference (NR, or blind). Full-reference quality metrics compare two images, while no-reference metrics assess the visual quality of a single image. Adversarial attacks on image quality metrics usually aim to modify a low-quality image so that a target quality assessment metric can predict its quality as high. As a result, the adversarial image is assigned a high-quality score despite being perceived by humans as low quality. The field of vulnerability analysis for novel NR metrics is rapidly growing \cite{bonnet2020fooling, yang2024exploring, leonenkova2024ti, kashkarov2024can, deng2024sparse, konstantinov2024image, yang2024beyond, zhang2024vulnerabilities, ran2025black, meftah2023evaluating, siniukov2023unveiling}. 

In \cite{korhonen2022adversarial}, the authors generate adversarial perturbations for NR metrics with perturbations hidden in textured regions via the Sobel filter. 
In \citep{Shumitskaya_2022_BMVC}, NR metrics were attacked using a universal adversarial perturbation (UAP) approach. 
In \citep{DBLP:conf/iclr/ShumitskayaAV23}, the authors proposed the FACPA attack, which uses an additional convolutional neural network to fasten the adversarial perturbations generation. 
In \cite{zhang2022perceptual}, a two-step perceptual attack for NR metrics was proposed. The authors 
generate a series of perturbed images with different levels of distortion visibility and then ask human observers to assess whether the perturbations in each image are perceptible. As a result, they collected adversarial images with invisible distortions that caused NR metrics to yield higher scores. 
In \cite{pmlr-v235-shumitskaya24a}, in contrast with the previous one, a method for adversarial attacks on NR metrics that creates invisible distortions without requiring human evaluation was proposed. 
In \cite{antsiferova2024comparing}, using a comprehensive set of adversarial attacks, the authors proposed a methodology for testing NR and FR IQA metrics' robustness and published an open benchmark. 

The field of IQA metrics' defense against adversarial attacks is much less developed. Several studies were dedicated to defenses for IQA metrics. Korhonen et al. \cite{korhonen2022adversarial} explored basic defense mechanisms for NR IQA metrics, such as adversarial attack detection, image preprocessing, and adversarial training. 
In \cite{SHUMITSKAYA2024103913}, the authors showed that resizing and random cropping help reduce the effectiveness of some attacks for NR metrics. Adversarial training was used in \cite{kettunen2019lpips, ghazanfari2023r} to increase the robustness of the LPIPS quality metric and modified in \cite{chistyakova2024increasing} to improve the performance of adversarially trained NR IQA metrics. 
In \cite{gushchin2024adversarial, gushchin2024guardians}, an application of adversarial purification methods to an IQA task was deeply explored. Liu et al. \cite{liu2024defense} showed that gradient norm regularization improved IQA metrics robustness.
Zhang et al. ~\cite{zhang2024secure} proposed a method of robust VQA using spatial grid sampling, pixel-wise randomization, and temporal information extraction.
However, these defenses are primarily empirical. Ghazanfari et al.~\cite{ghazanfarilipsim} proposed a provably robust perceptual similarity metric (LipSim) by leveraging 1-Lipschitz neural network as the backbone. This approach imposes certain restrictions on the metric's architecture, leading to a drop in model accuracy. Implementing similar architectures without compromising accuracy is challenging for NR metrics, as they typically have more sophisticated architectures. Gushchin et al.~\cite{gushchin2024guardians} tested existing randomized smoothing-based certified methods for NR IQA. However, a straightforward implementation of these methods reduced the correlation of these metrics. Our paper focuses on developing randomized smoothing-based approaches explicitly designed for certified NR IQA to avoid limiting the model architecture.

\subsection{Randomized Smoothing for Image Classification}
\textbf{Randomized smoothing (RS)} \citep{cohen2019certified} is a technique that transforms any classifier, which performs well under Gaussian noise, into a new classifier that is certifiably robust to adversarial perturbations of a bounded $l_2$ norm. Namely, if $f:\mathbb{R}^d \to \mathbb{R}$ is the base classifier, its randomized counterpart is defined as 
\begin{equation}
\label{eq:rs}
    g(x) = \mathbb{E}_{\varepsilon \sim \mathcal{N}(0,\sigma^2I)} f(x + \varepsilon).
\end{equation}
When $f(x)$ is the bounded function, the smoothed classifier in the form from Eq. \eqref{eq:rs} has a uniformly bounded gradient and, hence, is a Lipchitz function: one can prove that for all the points from a specific neighborhood of the input point $x$, it predicts the same class index as for  $x$. Although randomized smoothing is model-agnostic, it has an accuracy lower than the source model since its evaluation in a single point implies assessing the source classifier over its noisy neighborhood. 
To address this issue, \cite{salman2020denoised} extended RS by appending a denoiser network to the source classifier. This approach aims to train a denoiser that not only reconstructs the original image but also aligns the predictions of the smoothed model with the ones of the source model. The authors experimented with two architectures for the denoiser network, DnCNN \citep{zhang2017beyond} and MemNet \citep{tai2017memnet}, and demonstrated that the application of the denoiser significantly improves the accuracy of the smoothed classifier, in comparison to vanilla randomized smoothing. This approach was lately extended in \cite{carlini2022certified}, where the convolution-based denoiser if replaced with the pre-trained denoising diffusion probabilistic model. In \cite{chen2022densepure}, the authors proposed the \textbf{DensePure (DP)} method, which involves multiple runs of denoising through the reverse process of a diffusion model using different random seeds to generate multiple samples. 

\subsection{Median Smoothing for the Regression Models}

In \cite{chiang2020detection}, the authors proposed a method to certify regression models using \textbf{Median Smoothing (MS)}. In contrast to the randomized smoothing, the prediction for the sample $x$ is computed as the median of predictions over the input space. Namely, median smoothing has the form 
\begin{equation}
    g(x) = \text{med}_{\varepsilon \sim \mathcal{N}(0, \sigma^2I)} f(x+\varepsilon). 
\end{equation}
The authors demonstrated that the median smoothing yields more robust regression model than randomized smoothing, since the mean is more sensitive to outliers.  They have shown that extending median smoothing to \textbf{Denoised Median Smoothing (DMS)} --- by incorporating a denoising network before computing predictions on noisy samples --- improves accuracy in the object detection task.


%% file: sec/3_finalcopy.tex
\section{Proposed method}
In this section, we formulate the problem statement and provide a detailed description of the proposed method. 
\subsection{Problem statement}

Given the fixed image quality assessment metric $M: \mathbb{R}^d \to \mathbb{R}_{+},$ predefined constants $\delta > 0$ and $\varepsilon > 0$, the goal of the paper is to design an operator $G: \mathbb{R}_{+} \to \mathbb{R}_{+}$ that satisfies the property of local robustness to additive perturbations and does not significantly degrade the quality of IQA metric $M$. We formulate the robustness in terms of small deviations of $G$ under small-norm perturbations of the input of $M$. Namely, such an operator has to satisfy
    \begin{equation}
        \vert G(M(x)) - G(M(x + \Delta x)) \vert \le \delta \  \text{for all} \ \|\Delta x\|_2 \le \varepsilon,
    \end{equation}
    where $x \in  \mathbb{R}^d$ is the fixed image and $\Delta x$ is an additive perturbation. 
   To ensure that the operator does not decrease the quality of the IQA metric significantly, we introduce the concept of $\tau-$closeness.  

    \begin{definition}[$\tau-$closeness]
    \label{t_coeff}
    Let $S = \{x_1, \dots, x_m\}$ be the set of images and $Y = \{y_1, \dots, y_m\}$ be the set of corresponding subjective quality scores. Let $X_1 = (M(x_1), \dots, M(x_m))^\top$ and $X_2 = (G(M(x_1)), \dots, G(M(x_m)))^\top$. Then, the operator $G$ is said to be $\tau-$close on $S$, if 
    \begin{equation}
        \vert \rho(X_1, Y) - \rho(X_2, Y) \vert \le \tau,
    \end{equation}
    where $\tau > 0$ is the predefined threshold and $\rho(X,Y)$ is the  Pearson correlation coefficient.   

    \end{definition}
    \begin{remark}
        In this work, we evaluate the $\tau-$closeness in terms of both Pearson correlation and Spearman's rank correlation $r(X,Y) = \rho(R(X), R(Y))$, where $R(X)$ and $R(Y)$ are the rank variables. 
    \end{remark}
    In a nutshell, $\tau-$closeness reflects how $G$ affects the correlation between the metric's outputs and the subjective scores assigned to the input images. Note that, given $\varepsilon$, the tuple $(\delta, \tau)$ represents both the robustness and performance of  $G$.

\subsection{Method}



As the baseline for $G$, we choose the Median Smoothing proposed in \citep{chiang2020detection}. 

\subsubsection{Median Randomized Smoothing for IQA} 
Given the base image-quality metric $M: \mathbb{R}^d \rightarrow \mathbb{R}_{+}$, its Median Smoothed analog can be written as follows as 
\begin{equation}
G(M(x)) = \text{med}[M(x+r)], \text{ where } r \sim \mathcal{N}(0, \sigma^2 I ).
\label{eq:e2}
\end{equation}
Here $x$ is the input image, $I$ is an identity matrix, $r \sim \mathcal{N}(0, \sigma^2 I)$ is the sample of Gaussian noise and $\text{med}(\Omega)$ denotes the median of the finite set $\Omega$.  

According to \citep{chiang2020detection}, the deviation of the median smoothing in the form from \eqref{eq:e2} is bounded. Namely, the following theorem holds:
\begin{theorem}
Given $G(M(x))$ in the form from \eqref{eq:e2}, for all $\|u\|_2 \le \varepsilon,$
\begin{equation}
\underline{H}_{\underline{p}}(M(x)) \leq G(M(x+u)) \leq \overline{H}_{\overline{p}}(M(x)).
\label{eq:theorem1}
\end{equation}
Here, $\underline{p} = \Phi(- \frac{\varepsilon}{\sigma})$, $\overline{p} = \Phi(\frac{\varepsilon}{\sigma}) $ and $\Phi(\cdot)$ is the Gaussian cumulative density function.  $\underline{H}_{\underline{p}}(M(x))$ and $\overline{H}_{\overline{p}}(M(x))$ are defined as the percentiles of the smoothed function:
\begin{align}
    &\underline{H}_{\underline{p}}(M(x)) = \text{sup}_{y \in \mathbb{R}}\{\mathbb{P}_{r \sim \mathcal{N}(0, \sigma^2 I)}[M(x+r) \leq y] \leq \underline{p} \}, \nonumber \\ 
    &\overline{H}_{\overline{p}}(M(x)) = \text{inf}_{y \in \mathbb{R}} \{\mathbb{P}_{r \sim \mathcal{N}(0, \sigma^2 I)} [M(x+r) \leq y] \geq \overline{p} \}.  \nonumber 
\end{align}
\end{theorem}

\paragraph{Auxiliary image denoiser.}
It is noteworthy that the straightforward application of the median smoothing to an image quality assessment metric will lead to a significant performance degradation of the latter: the variance of the smoothed IQA metric decreases, and so does the correlation between the metric's output and the subjective scores \cite{ohayon2023reasons}. To solve this problem, we propose to train an auxiliary image denoiser as a part of the smoothed image quality assessment pipeline: the denoiser is embedded after the input images are convolved with the noise but before the noised images are passed to the IQA metric. 

For the denoiser training, we propose a loss function composed of three components: $MSE_{loss}$, $RANK_{loss}$, and $TARG_{loss}$.

To define the components of the loss function, we firstly introduce the notation: let $D:\mathbb{R}^d \to \mathbb{R}^d$ be the denoiser parameterized by a neural network, $B = \{(x_i, y_i)\}_{i=1}^N$ be the batch of input images, $\tilde{B} = \{(x_i + r_i, y_i)\}_{i=1}^N$ be the batch of images with the noise, where $r_i \sim \mathcal{N}(0, \sigma^2 I)$ from the Eq. \eqref{eq:e2} and let the images be of dimension $3 \times d_1 \times d_2.$ Then, the components of the loss function are defined as follows.

\paragraph{Mean squared error.} The first component of the loss function is the mean squared error between the pixels of the original images and the denoised samples. Optimization of this component during the training phase forces the denoiser to remove the noise added to the input images effectively.  Specifically, the MSE term is given as follows:
\begin{equation}
    MSE_{loss} (B, \tilde{B})=  \frac{1}{3  N  d_1  d_2} \sum_{i=1}^N \|x_i - D(\tilde{x}_i)\|^2_2,
\end{equation}
where $x_i$ and $\tilde{x}_i$ are the $i'$th images from $B$ and $\tilde{B},$ respectively.

\paragraph{Ranking loss.} To ensure that the smoothed metric does rank the images within the training batch in the same manner as the target subjective scores, we introduce the ranking loss term. We assume that, during training, the IQA metric is the composition $M(D(\cdot))$ and the number of noise samples applied to each image equals to one. Let $p = M(D(\tilde{B}))$ be the vector of objective scores assigned by the smoothed IQA metric to the images from $\tilde{B}$ and $t = t(B)$ be the vector of subjective scores assigned to original images from $B$.  Then, we can introduce the ranking matrix  element-wise:
\begin{equation}
    R(B, \tilde{B})_{i,j} =  (p_i - p_j)  \text{sign}(t_j - t_i).
\end{equation}
Intuitively, if the $i'$th and $j'$th images from $B$ are ranked by target subjective scores and smoothed IQA metric differently, the $ij'$th term of the ranking matrix penalizes this inconsistency. 
Thus, the ranking loss term is defined as follows:
\begin{equation}
    RANK_{loss} (B, \tilde{B}) = \frac{1}{N^2}\frac{\sum_{i,j}\max(0, R(B, \tilde{B})_{i,j})}{1  +  \| R(B, \tilde{B})\|_\infty}.
\end{equation}

\paragraph{Target loss.} We introduce the target loss to ensure that the predictions of the original IQA metric and its smoothed counterpart on the clean images do not differ much. Optimizing the target loss ensures the consistency and effectiveness of the smoothed IQA metric when applied to the clean data. The target loss term is given as follows:
\begin{equation}
    TARG_{loss} (B, \tilde{B}) = \frac{1}{N} \sum_{i=1}^N [M(x_i) - M(D(\tilde{x}_i))]^2.
\end{equation}

Overall, the training loss function is a linear combination of the three terms above:
\begin{align}
\label{eq:denoiser_loss}
    L(B, \tilde{B}) = & MSE_{loss}(B, \tilde{B}) + C_r RANK_{loss}(B, \tilde{B}) + \nonumber \\ &C_t TARG_{loss}(B, \tilde{B}).
\end{align}
Minimization of this combined loss forces the denoiser to balance between accurately removing the noise and ensuring that the smoothed IQA metric's predictions are aligned with the source metric. In the Supplementary materials, we provide the results of experiments on finding the optimal values of $C_r$ and $C_t$ from Eq. \eqref{eq:denoiser_loss} and the batch size $N$.

\paragraph{Denoiser architecture.} To ensure the computational efficiency of the smoothed IQA metric, we use lightweight U-Net \cite{ronneberger2015u} with $7$ layers as the denoiser. Other more complex architectures were tested, like DnCNN \cite{zuo2018convolutional} and diffusion-based ones. However, they were $10-100$ times slower than the U-Net while achieving almost the same performance in terms of correlations between the predictions of the smoothed IQA metric and the subjective scores.

\begin{remark}
    Together with the auxiliary image denoiser $D$, the smoothed image quality assessment metric is defined as the composition $G(M(D(\cdot)))$ in the form
    \begin{align}
        & G(M(D(x))) =  \normalfont \text{med}[M(D(x+r))], 
    \end{align}
    {where $r \sim \mathcal{N}(0, \sigma^2I)$}.
\end{remark}

\subsection{Bounded Deviation of Rank Correlation}
In this section, we discuss the connection between the deviation of the smoothed IQA metric from the source IQA metric and the deviation of rank correlation. Suppose that the deviation of the smoothed metric from the source metric is bounded; namely, let $X = \{x_1, \dots, x_m\}$ be the set of data points and 
\begin{equation}
    \Delta(X) = \|G(M(D(X))) - M(X)\|_\infty
\end{equation}
be the deviation of the smoothed metric from the source metric on $X$. Then, the following lemma holds.
\begin{lemma}
    Let $a_i = G(M(D(x_i))), b_i = M(x_i)$. 
    Suppose that  $\delta^{ij} = |b_{i} - b_{j}|$ and let the sequence $B = \{\delta^{ij}: i \ne j\}$ be sorted in a non-descending order:
    \begin{equation}
       B = \{\delta_{k_1} \le \delta_{k_2} \le \dots \le \delta_{k_{m(m-1)/2}}\}
    \end{equation}
    Note that the smoothed metric $G(M(D(\cdot)))$ makes a ranking error on the pair $(x_i, x_j): i \ne j$  of samples from $X$ if the ranking variables are related as
    \begin{equation}
       [R(a_i) - R(a_j)] [R(b_i) - R(b_j)] < 0.  
    \end{equation}
    Hence, if 
    \begin{equation}
        t = \arg\min\limits_{\hat{\delta} \in B}: \hat{\delta}> \Delta(X),
    \end{equation}
    then $G(M(D(\cdot)))$ makes at most $t$ out of $m(m-1)/2$ possible ranking errors on samples from $X$. 
\end{lemma}




\section{Experiments}

We compared our method to the previous approaches (MS and DMS \cite{chiang2020detection}), assessing five learning-based NR IQA metrics on three datasets of images.  

\paragraph{Datasets.} We compared IQA methods on three datasets: KonIQ \cite{hosu2020koniq}, CLIVE \cite{ghadiyaram2015massive} and SPAQ \cite{fang2020perceptual}. These datasets contain subjectively assessed images that cover a broad range of resolutions, aspect ratios, and real-world distortions captured by a camera. Each dataset was divided into train ($80\%$ images), validation ($10\%$ images), and test ($10\%$ images) subsets. Table \ref{tab:datasets} contains a detailed description of these datasets.

\begin{table}[htb]
\caption{Description of the datasets used for comparison. DSLR stays for a Digital single-lens reflex camera. DSC stays for the Digital still camera.}
\label{tab:datasets}
\begin{center}
\begin{small}
\begin{tabular}{lccc}
\toprule
 Dataset & \# images & Type of camera & Subj. env. \\
\midrule
KonIQ \cite{hosu2020koniq} & 10,073 & \makecell{DSLR/DSC/\\Smartphone} & Crowdsourcing  \\
CLIVE \cite{ghadiyaram2015massive} & 1,162 & \makecell{DSLR/DSC/\\Smartphone} & Crowdsourcing  \\
SPAQ \cite{fang2020perceptual} & 11,125 & Smartphone & Laboratory  \\
\bottomrule
\end{tabular}
\end{small}
\end{center}
\end{table}
\paragraph{IQA metrics.} For experiments, we selected KonCept \cite{hosu2020koniq}, Hyper-IQA \cite{su2020blindly}, CLIP-IQA+ \cite{wang2023exploring}, DBCNN \cite{8576582} and Topiq \cite{chen2024topiq} NR IQA metrics. These metrics were chosen to cover different architectures as shown in Table \ref{tab:iqa_models}.

\begin{table}[htb]
\caption{Description of the IQA metrics used for experiments.}
\label{tab:iqa_models}
\begin{center}
\begin{small}
\begin{tabular}{lccc}
\toprule
IQA metric & Backbone & \# params  \\
\midrule
KonCept \cite{hosu2020koniq} & InceptionResNetV2 & 59.82M \\ 
Hyper-IQA \cite{su2020blindly} & ResNet50 & 27.38M \\
CLIP-IQA+ \cite{wang2023exploring} & CLIP & 150M \\
DBCNN \cite{8576582} & S-CNN & 15.31M \\
Topiq \cite{chen2024topiq} & CFANet & 45.20M \\
\bottomrule
\end{tabular}
\end{small}
\end{center}
\end{table}
\paragraph{Use cases.} We set the number of noise  samples for smoothing to $2,000$ to balance between the computational efficiency of the method and estimation accuracy. Recall that $\overline{p} = \Phi (\frac{\epsilon}{\sigma})$, where $\Phi$ is the inverse  CDF of the standard Gaussian random variable. To ensure the method's applicability, the number of samples $N$ must satisfy $\ceil{\overline{p}*N} < N$, where $\ceil{\cdot}$ denotes the ceiling function. That is, when $N$ is set to 2,000, $\overline{p} < 0.9975$ and $\frac{\epsilon}{\sigma} < 3.295$, limiting the range of epsilons we can consider for a given $\sigma$. Also, it is better to avoid using extreme values for $\frac{\epsilon}{\sigma}$ as this can significantly increase the certified delta due to the impact of outliers. We considered two use cases: $\sigma=0.12, \epsilon=0.06$ ($\frac{\epsilon}{\sigma}=0.5$) and $\sigma=0.18, \epsilon=0.36$ ($\frac{\epsilon}{\sigma}=2$). These use cases were chosen to estimate the method performance to defend against attacks with adversarial perturbation $u$ of two strengths: $||u||_2 <= 0.06$ (weak) and $||u||_2 <= 0.36$ (strong). These parameters were selected to ensure strong defended metric performance in terms of SROCC and tight certified guarantees. Higher values of $\sigma$ result in lower defended metric accuracy, as indicated by a decrease in SROCC. Values of $\sigma$ above $0.18$ lead to poor correlations, even with smoothing combined with denoising, due to significant degradation during the noise application step. 

\subsection{Comparison on clean images}
For DMS and DMS-IQA methods, we trained denoisers on $80\%$ of images from each dataset. For the DMS method, a single denoiser for each use case ($\sigma=0.12$ and $\sigma=0.18$) was trained for all IQA metrics, as this approach does not require model-specific knowledge during training. Training was conducted for $30$ epochs with the learning rate of $lr=0.001$. For DMS-IQA method, we trained a separate denoiser for each use case and each base IQA metric. Training was conducted as a fine-tuning from the DMS weights for $50$ epochs with the learning rate of $lr=0.0001$. The models best on validation ($10 \%$ of images) were used for testing. We tested the aforementioned methods on three test subsets drawn from different datasets: KonIQ ($10\%$ of images), SPAQ ($10\%$ of images), and CLIVE ($10\%$ of images). To evaluate the performance of defense methods, we computed $\tau_{SROCC}$ and $\tau_{PLCC}$ scores (in the form from Eq. \ref{t_coeff}) between the defended metric's predictions and actual subjective scores, provided by human observers. We also measured certified deltas to evaluate certified guarantees of compared methods. Certified Delta (CD) is the difference $CD = S^u - S^l$ of upper and lower bounds, which certified methods guarantee on some image $x$ for any perturbation $\Delta x$ with an $l_2$ norm less than $\epsilon$. We divided these deltas by the IQA metric range to ensure comparability of deltas across metrics. All results reported in the paper contain certified deltas as a percentage relative to the metric range.

\subsection{Comparison on adversarial images}
\paragraph{Adversarial examples generation.}
To conduct an adversarial attack, we selected $100$ random images from the KonIQ dataset and generated adversarial examples for them using the method proposed in  \cite{goodfellow2014explaining}, where L-BFGS is used to solve a constrained optimization problem. 
Namely, we solve the following optimization problem:
\begin{align}
\label{eq:e4}
& \max\limits_{\Delta x} [M(x + \Delta x) - M(x)] \\ & \text{ s.t.} \|\Delta x\|_2 \leq \epsilon. \nonumber
\end{align}
In our case, we use the backpropagation method instead of L-BFGS for efficiency reasons. Specifically, we perform $1000$ backpropagation steps to minimize the loss function $\hat{l}$ using the Adam optimizer with a learning rate of $lr=0.0005$.  We generated adversarial examples for $\epsilon=0.06$ and $\epsilon=0.36$ $l_2$ norms for each tested IQA NR metric. The loss function $\hat{l}$ is given in the equation below:
\begin{equation}
\begin{array}{cc}
\hat{l}(\Delta x) = \frac{M(x) - M(x + \Delta x)}{\text{range}(M)} + \max(0, \|\Delta x\|_2 - \epsilon).
\end{array}
\label{eq:e5}
\end{equation}
In Eq. \eqref{eq:e5}, $\text{range}(M)$ corresponds to the diameter of the set of values of $M$, namely, 
\begin{equation}
\text{range}(M) = \sup_{x,y \in \mathbb{R}^d} |M(x) - M(y)|.
\end{equation}

\paragraph{Evaluation on adversarial examples.} When adversarial examples are generated, we compared the performance of the undefended IQA metric with the performance of the IQA metric defended using MS, DMS, and DMS-IQA techniques. 
Let $x$ and $x_{adv}$ represent clean and adversarial images and $(S, S^l, S^u)$, $(S_{adv}, S^l_{adv}, S^u_{adv})$ denote the answers of certified IQA metric for these images. We define the adversarial gain as $Adv.Gain = (S_{adv} - S)/range(M)$ and certified bounds $CD^u$ and $CD^s$ as differences $CD^u = (S^u - S)/range(M)$ and $CD^l = (S - S^l)/range(M)$, where $range(M)$ is the range of base IQA metric scores.

\section{Results}
\subsection{Results on clean images}
Table \ref{tab:results_clear_data} compares the proposed DMS-IQA certified defense with two prior methods (MS and DMS \cite{chiang2020detection}). The comparison in made on three datasets (KonIQ \cite{hosu2020koniq}, CLIVE \cite{ghadiyaram2015massive} and SPAQ \cite{fang2020perceptual}) and for two use cases: weak ($\sigma=0.12, \epsilon=0.06$) and strong ($\sigma=0.18, \epsilon=0.36$). We observe that DMS demonstrates a notable improvement over MS in terms of $\tau_{SRCC}$ and $\tau_{PLCC}$ scores, indicating that image denoising with MS enhances metric accuracy. We also can see that the proposed DMS-IQA method achieves better $\tau_{SRCC}$ and $\tau_{PLCC}$ scores while maintaining certified deltas comparable to those of the MS and DMS methods for both use cases. This suggests that the denoiser trained using the technique described in this paper offers a more effective method for defending NR IQA metrics.

\begin{table*}[htb]
\caption{The results of comparison proposed DMS-IQA method with two prior methods --- MS and DMS \cite{chiang2020detection} --- on three datasets for NR IQA: KonIQ \cite{hosu2020koniq}, CLIVE \cite{ghadiyaram2015massive} and SPAQ \cite{fang2020perceptual}. The results are averaged across five NR-IQA metrics: KonCept \cite{hosu2020koniq}, Hyper-IQA \cite{su2020blindly}, CLIP-IQA+ \cite{wang2023exploring}, DBCNN \cite{8576582} and Topiq \cite{chen2024topiq}. Weak use case (w): ($\sigma=0.12, \epsilon=0.06$), strong use case (s): ($\sigma=0.18, \epsilon=0.36$). Certified Delta (CD) is expressed as a percentage relative to the metric range.}
\label{tab:results_clear_data}
\begin{center}
\begin{small}
\begin{tabular}{lccc|ccc|ccc}
\toprule
 & \multicolumn{3}{c}{KonIQ \cite{hosu2020koniq}} & \multicolumn{3}{c}{CLIVE \cite{ghadiyaram2015massive}} & \multicolumn{3}{c}{SPAQ \cite{fang2020perceptual}} \\
Method & $\tau_{SROCC} \downarrow$ &  $\tau_{PLCC} \downarrow$ &  $CD, \% \downarrow$ & $\tau_{SROCC} \downarrow$ &  $\tau_{PLCC} \downarrow$ &  $CD, \% \downarrow$ & $\tau_{SROCC} \downarrow$ &  $\tau_{PLCC} \downarrow$ &  $CD, \% \downarrow$ \\
\midrule
MS (w) & $0.2542$ & $0.2949$ & $\underline{1.3438}$ & $0.2911$ & $0.3176$ & $\textbf{1.4378}$ & $0.1279$ & $0.1292$ & $\textbf{1.2985}$ \\
DMS (w) & $\underline{0.1015}$ & $\underline{0.0899}$ & $1.5241$ & $\underline{0.0514}$ & $\underline{0.0517}$ & $1.8877$ & $\underline{0.0160}$ & $\textbf{0.0083}$ & $1.7374$ \\
DMS-IQA (w) & $\textbf{0.0612}$ & $\textbf{0.0750}$ & $\textbf{0.9646}$ & $\textbf{0.0216}$ & $\textbf{0.0388}$ & $\underline{1.6041}$ & $\textbf{0.0144}$ & $\underline{0.0241}$ & $\underline{1.4830}$ \\
\midrule
MS (s) & $0.3835$ & $0.4340$ & $\underline{5.2898}$ & $0.4338$ & $0.4611$ & $\textbf{5.5468}$ & $0.2543$ & $0.2563$ & $\textbf{4.8508}$ \\
DMS (s) & $\underline{0.1427}$ & $\underline{0.1289}$ & $6.8368$ & $\underline{0.0642}$ & $\underline{0.0694}$ & $8.2978$ & $\underline{0.0355}$ & $\textbf{0.0239}$ & $7.6146$ \\
DMS-IQA (s) & $\textbf{0.0851}$ & $\textbf{0.1022}$ & $\textbf{4.2925}$ & $\textbf{0.0373}$ & $\textbf{0.0540}$ & $\underline{7.0629}$ & $\textbf{0.0230}$ & $\underline{0.0352}$ & $\underline{6.2755} $\\
\bottomrule
\end{tabular}
\end{small}
\end{center}
\end{table*}

\subsection{Results on adversarial images}

\begin{figure}[ht]
\begin{center}
\centerline{\includegraphics[width=\linewidth]{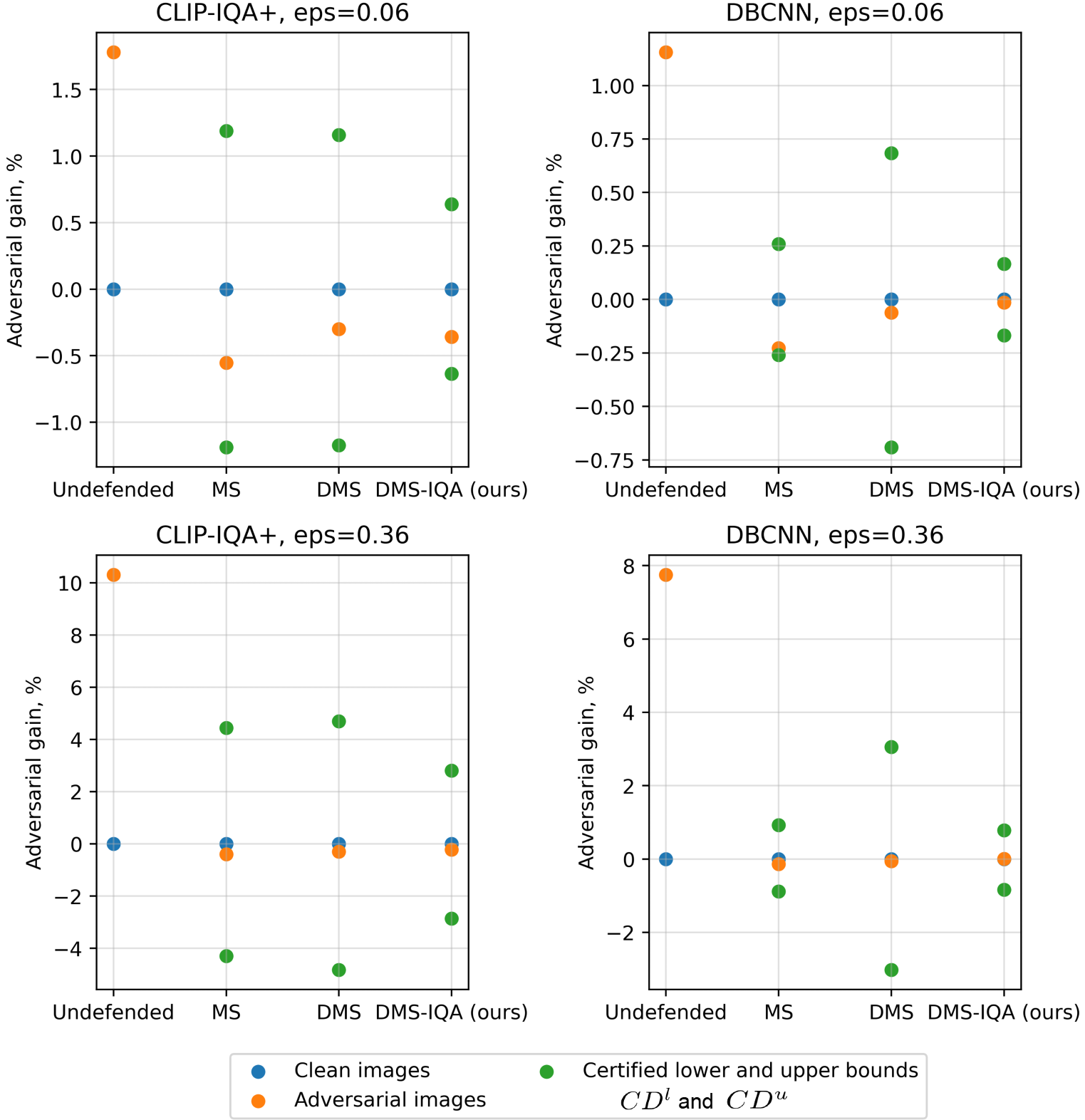}}
\caption{Comparison of the proposed DMS-IQA method with MS, DMS and undefended metric on adversarial images for CLIP-IQA+ and DBCNN NR IQA metrics. Green dots represent the certified guarantees. Results on the plot are averaged across images.}
\label{fig:adv_cert_examples}
\end{center}
\end{figure}

Figure \ref{fig:adv_cert_examples} presents the results of experiments for CLIP-IQA+ and DBCNN NR IQA metrics for two use cases: $\epsilon = 0.06$ and $\epsilon = 0.36$. We calculated the percentage of metric change (adversarial gain) after the attack, as well as certified bounds on clean images. For certified defenses --- MS, DMS, and DMS-IQA --- the metric score after the attack (orange dot) consistently falls within the certified lower and upper bounds (green dots). For the undefended metric, certified bounds are undefined. Additionally, we observe that the change in metric score for the undefended metric after the attack is greater than half of the certified delta. In other words, it exceeds the absolute value of the upper bound relative to the clean metric score. Table \ref{tab:results_adv_data} contains the results of the same experiment for all five NR IQA metrics. 

\begin{table}[htb]
\caption{Comparison of the proposed DMS-IQA method with MS, DMS and undefended metric on adversarial images for five NR IQA metrics: KonCept \cite{hosu2020koniq}, Hyper-IQA \cite{su2020blindly}, CLIP-IQA+ \cite{wang2023exploring}, DBCNN \cite{8576582} and Topiq \cite{chen2024topiq}. Results are averaged across images and NR IQA metrics.}
\label{tab:results_adv_data}
\begin{center}
\begin{small}
\begin{tabular}{llccc}
\toprule
$\epsilon$ & Method & Adv.Gain$\downarrow$ & $CD^l\uparrow$ & $CD^u\downarrow$   \\
\midrule
\multirow{4}*{$0.06$} & No-Def. & $0.865$ & $-\infty$ & $+\infty$ \\
& MS & $0.418$ & $\underline{-0.682}$ & $\underline{0.684}$ \\
& DMS & $\underline{0.307}$ & $-0.794$ & $0.789$ \\
& DMS-IQA & $\textbf{0.251}$ & $\textbf{--0.495}$ & $\textbf{0.493}$ \\
\midrule
\multirow{4}*{$0.36$} & No-Def. & $5.038$ & $-\infty$ & $+\infty$ \\
& MS & $\underline{0.261}$ & $\underline{-2.583}$ & $\underline{2.680}$ \\
& DMS & $0.266$ & $-3.482$ & $3.418$ \\
& DMS-IQA & $\textbf{0.144}$ & $\textbf{--2.168}$ & $\textbf{2.083}$ \\
\bottomrule
\end{tabular}
\end{small}
\end{center}
\end{table}


\section{Discussion}
\paragraph{Usage in the loss function.} NR metrics are often used in loss functions for training image- and video- processing algorithms, however, due to vulnerability to adversarial attacks, it can cause appearance of some artifacts \cite{kashkarov2024can}. We did additional experiments regarding the usage of smoothed NR IQA metric in the loss function on the example of image denoising. As a base metric, we selected the fastest NR IQA metric used for testing --- DBCNN. For 100 images, we generated noisy versions by adding Gaussian noise with $\sigma=0.1$.  This produced adversarial examples with a mean RMSE of 0.2. Next, we iteratively denoised these images using the following loss function: $loss = 1 - Q(x, y)/100 + MSE(x, y) / 1000$, where $Q$ is a quality metric: undefended DBCNN, DBCNN+MS, DBCNN+DMS or DBCNN+DMS-IQA. The division by 100 is applied because the score range of the DBCNN metric is approximately 100. Also, we divided the MSE component by 1000 to reduce its significance. Since certified restrictions for methods work for very low $l_2$ norm, it is important to set a minimal learning rate. We used a learning rate $lr=0.00001$ and performed 1000 iteration steps for each image. In this experiment, we used the number of samples $N=100$ for all certified methods. Table \ref{tab:opt_exp} contains the results of this experiment. Optimizing the IQA metric smoothed by simple MS does not improve the resulting RMSE. We assume this is because the performance of the DBCNN+MS metric, in terms of $\tau_{SROCC}$ and $\tau_{PLCC}$ scores, is lower than that of the other metrics. The DMS-IQA method showed the best results in this experiment, suggesting that using a smoothed metric for optimizing image processing algorithms may be reasonable. Smaller samples (10–100) can be used for better efficiency.

\begin{table}[htb]
\caption{RMSE after optimization of original and median-smoothed DBCNN \cite{8576582} metrics from noisy images. Results averaged over $100$ images. }
\label{tab:opt_exp}
\begin{center}
\begin{small}
\begin{tabular}{lc}
\toprule
NR IQA metric & RMSE $\downarrow$ \\
\midrule
DBCNN & $0.17316$ \\
DBCNN+MS & $0.17574$ \\
DBCNN+DMS & $\underline{0.17041}$ \\
DBCNN+DMS-IQA & $\textbf{0.15924}$ \\
\bottomrule
\end{tabular}
\end{small}
\end{center}
\end{table}

\paragraph{Computational complexity.} 
We measured the average time in seconds spent for assessing one image of size $512 \times 384$ for DMS-IQA defended and undefended NR IQA metrics. For certified defense, the number of samples was $2000$. We observe that certified defense increases the processing time by approximately $300$. Calculations were made on 7 NVIDIA Tesla A100 80 Gb GPU, Intel Xeon Processor 354 (Ice Lake) 32-Core Processor @ 2.60 GHz. The proposed method demands huge computational resources to evaluate high-resolution data. However, with the continuous advancement of computational power, it's likely that in a few years, high-performance computing may be accessible even on personal devices. This potential future accessibility makes median smoothing methods a promising direction for developing certified IQA methods.



\paragraph{Multi-dataset training.}
We observe the poor transferability of a method trained on a single dataset to other datasets. Because of that, we conducted an additional experiment involving multi-dataset training. Specifically, we trained a denoiser for DMS-IQA on three datasets, KonIQ \cite{hosu2020koniq}, CLIVE \cite{ghadiyaram2015massive}, and SPAQ \cite{fang2020perceptual}, and tested the model on $500$ randomly selected images from the KADID dataset \cite{lin2019kadid}. We chose the CLIP-IQA \cite{wang2023exploring} metric as a base IQA metric, as it has the highest SROCC on this dataset. We used the following parameters: $\sigma=0.12, \epsilon=0.06, N=2000$. The Table \ref{tab:multi-dataset-training} shows the results of this experiment. According to this experiment, multi-training boosts the method's transferability on unseen data.

\begin{table}[htb]
\caption{The results of comparison DMS-IQA trained on single dataset vs. DMS-IQA trained on multiple datasets}
\label{tab:multi-dataset-training}
\begin{center}
\begin{small}
\begin{tabular}{lccc}
\toprule
 & $\tau_{SRCC} \downarrow$ & $\tau_{PLCC} \downarrow$ & $CD, \% \downarrow$ \\
\midrule
DMS-IQA$_{KONIQ}$ & $0.10$ & $\underline{0.03}$ & $\underline{1.27}$ \\
DMS-IQA$_{CLIVE}$ & $\underline{0.07}$ & $\textbf{0.01}$ & $1.77$ \\
DMS-IQA$_{SPAQ}$ & $0.13$ & $0.06$ & $\textbf{1.21}$ \\
DMS-IQA$_{multi-train}$ & $\textbf{0.04}$ & $\textbf{0.01}$ & $1.84$ \\
\bottomrule
\end{tabular}
\end{small}
\end{center}
\end{table}

\section{Conclusion}
In this paper, we propose DMS-IQA, a novel approach to certify image quality assessment methods against additive perturbations of a bounded magnitude. The method is based on median randomized smoothing and leverages an auxiliary denoising network, yielding a robust IQA algorithm aligned in predictions with the source metric. Extensive experimental evaluation of the proposed method shows that DMS-IQA outperforms previous approaches in terms of correlation with the subjective scores while achieving a comparable robustness to additive perturbations. We also experimentally show that  embedding of the smoothed image quality assessment metric in the training process of image processing algorithms can be beneficial for the latter, although require additional computational resources. 